# Dynamics of ring polymer melts: Memory function approach.

*Nail Fatkullin, Carlos Mattea, Kevin Lindt , Siegfried Stapf and Margarita Kruteva*

N. Fatkullin, C. Mattea, K. Lindt, S. Stapf

Technische Physik II/Polymerphysik, Technische Universität Ilmenau, D-98684 Ilmenau, Germany

M. Kruteva

Jülich Centre for Neutron Science (JCNS-1), Forschungszentrum Jülich GmbH, 52428 Jülich, Germany

Corresponding author email: ntakull@gmail.com

We investigated the static and dynamic properties of a Rouse ring polymer modified by introducing an effective, spherically symmetric, attractive potential of entropic nature and a memory function describing the effect of dynamic entanglement. Renormalized Rouse formalism is used to approximate the time dependence of the memory matrix. The results obtained are in good agreement with existing experimental data and computer simulation results for various scaling exponents of the ring polymer at $Z \leq 60$ , $Z = N / N_e$ , where $N_e$ is the number of Kuhn segments between adjacent crosslinks in linear polymer melts, and is the number of Kuhn segments. For large molecular weights, a refined self-consistent approximation is proposed to describe the time dependence of the memory function. It has been shown that this approximation potentially allows one to describe the effect of dynamic localization, i.e. the decrease in the self-diffusion coefficient of the rings at a faster rate than any polynomial law as the molecular mass increases.

## 1. Introduction.

In recent years, there has been growing interest in studying the properties of polymer systems containing cyclic macromolecules (See, for example, [1-16] and literature cited therein). It may seem at first glance that, at high molecular weights, the differences in the physical properties of cyclic and linear polymer systems should disappear, since the microscopic interaction potentials are identical in these cases, and the influence of boundary conditions should decrease as the molecular weight increases. However, the reality is far from that simple. For example, in melts, as is well known, the distribution of conformations of linear macromolecules can be considered as ideal to a large extent (see, for example [17-21]). This means that the square of the radius of gyration of a linear macromolecule depends on its size as $R_{g,lin}^2 = \frac{Nb^2}{6}$, where $N$ is the number of Kuhn segments in the macromolecule, $b$ is the length of the Kuhn segment. If the behavior of cyclic macromolecules were ideal, the square of the radius of gyration of cyclic macromolecules in melts were expected to depend on the size as $R_g^2 = \frac{Nb^2}{12}$. However, experiments using neutron scattering and computer analysis show that the dependence on molecular weight is, in fact, weaker, i.e. $R_g^2 \propto b^2 N^{\nu}$ with an exponent $\nu$ approaching 2/3 for very large N>>1 . The difference in the linear dimensions of ring polymers between ideal and real melts is explained by so-called topological interactions. The sets of possible conformations in ideal and real ring polymer melts differ significantly. All conformations existing in real ring melts are also present in ideal ring melts; however, ideal ring melts may contain conformations that are not present in real ring melts. This leads to differences in the conformational distribution functions and characteristic linear dimensions measured experimentally. As a consequence, the effective interaction potential of an ideal ring deviates from that of real rings, which are predominantly of an entropic nature. In fact topological interactions in ring polymers represent a special case of entropic interactions, demonstrating collective many-body nature. Thus, due to the well-known Bogoliubov–Born–Green–Kirkwood–Yvon (BBGKY) closed-path problem, the dynamical equations become unsolvable and require approximations.

In our recent work [4], we proposed an approach in which the influence of topological interactions on the equilibrium static conformations of ring polymers in melts is accounted for using a central harmonic potential. The dynamic properties of the proposed model were described in the simplest possible way, namely using the classical Rouse model, which does not account for the so-called entanglement effect. Entanglements, in a sense, can be regarded as temporal topological interactions and it was shown that already this very simplified model has many interesting nontrivial properties, such as e.g. a quasi-plateau at increasing molecular mass for the time dependence of segmental mean squared displacement. In this paper, we investigate a model that accounts for the influence of entanglement effects on the dynamics of ring melts based on the Mori-Zwanzig memory function formalism.

## 2. Theoretical part

### 2.1. General considerations

Our approach is based on the formalism originally introduced by Zwanzig and Bixon [22, 23], which was then applied to polymer dynamics by Schweizer [24, 25] and others (see [26–28] and the references cited therein). The Zwanzig-Mori projection operator technique allows the Generalized Langevin Equation to be derived for the probe chain (see [24-30] and detailed discussion in supporting material [31], [30] contains a correction to the numerical coefficient of the memory function for Rouse normal modes of linear chains, which was used in earlier works). For the case of polymer systems, it has the following structure:

$$\begin{aligned}&\frac{d}{dt}\vec{p}_n(t) = -\frac{\partial}{\partial \vec{r}_n} W\left(\left\{\vec{r}_i(t)\right\}\right) - \sum_k \int_0^t \Gamma_{nk}^{\alpha\beta}(\tau;t-\tau) v_k^{\beta}(t-\tau)\vec{e}_{\alpha} d\tau_{\alpha} + \vec{F}_n^{Q}(t) \\ &\frac{d}{dt}\vec{r}_n(t) \equiv \vec{v}_n\left(t\right) = \frac{1}{m}\vec{p}_n(t)\end{aligned} , \qquad (1)$$

where $\vec{p}_n(t)$ is the momentum of the polymer segment with number *n* at time moment *t*, $\vec{r}_n(t)$ is its position vector at time moment *t*, $v_n^{\beta}(t-\tau)$ is the $\beta$ component of its velocity at time moment $t-\tau$, $\vec{e}_{\alpha}$ is the unit vector aligned along the $\alpha = x, y, z$ axis, $\vec{F}_n^{Q}(t)$ is the so-called Generalized Langevin force by which the matrix acts on the segment with number *n*, $\Gamma_{nk}^{\alpha\beta}(\tau;t-\tau)$ is the so-called the memory matrix, *m* is the mass of the segment, and $W\left(\left\{\vec{r}_i(t)\right\}\right)$ is the effective intramolecular potential or mean-field potential. For times comparable to, and longer than, the segmental relaxation time $t \geq \tau_s = 10^{-11} \div 10^{-9}\, s$, the left-hand side term in Eq. (1) becomes negligible and the system of 6N stochastic integral-differential relations is reduced to a system of 3N stochastic integral-differential relations:

$$\frac{\partial}{\partial \vec{r}_n} W\left(\left\{\vec{r}_i(t)\right\}\right) = -\sum_k \int_0^t \Gamma_{nk}^{\alpha\beta}(\tau; t-\tau) v_k^{\beta}(t-\tau) \vec{e}_{\alpha} d\tau + \vec{F}_n^{Q}(t), \qquad (2)$$

where summation is assumed over repeating indices $\alpha$ , $\beta$ and $k$ .

The memory matrix $\Gamma_{nk}^{\alpha\beta}(\tau;t-\tau)$ is the most complex object in expressions (1) and (2). It describes the dynamical entanglement effects in the system and a formally exact relation for it has the following form:

$$\begin{aligned}\Gamma_{nk}^{\alpha\beta}(\tau;t-\tau) &= \frac{1}{k_B T \rho_N^{*}(t-\tau)} \left\langle F_k^{Q\beta}(\gamma) \delta\left(\gamma_N - \gamma_N(t-\tau)\right) F_n^{Q\alpha}(\tau) \right\rangle \\ &= \frac{1}{k_B T} \left\langle F_k^{Q\beta}(0) F_n^{Q\alpha}(\tau) \right\rangle_{\gamma_N(t-\tau)}^{*}\end{aligned}, \qquad (3)$$

where

$$\gamma_N \equiv \left\{\vec{r}_1, \vec{r}_2, \ldots \vec{r}_N, \vec{p}_1, \vec{p}_2, \ldots \vec{p}_N\right\} \qquad (4)$$

is the set of all phase variables of the probe macromolecule, $\delta\left(\gamma_N - \gamma_N(t-\tau)\right)$ is the 6N dimensional Dirac $\delta$ function, $\rho_N^{*}(t-\tau)$ is the equilibrium distribution of phase variables of matrix under the condition, that phase variables of the probe macromolecule are fixed and they had at time moment $t-\tau$ . The time evolution of the Langevin stochastic forces $\vec{F}_n^{Q}\left(t\right)$ acting on the segments of the probe chain is described by what is called projection dynamics:

$$\vec{F}_n^{Q}\left(t\right) = \exp\left\{i\hat{Q}_N \hat{L} \hat{Q}_N t\right\} \delta\vec{F}_n^{\mathrm{inter}} \qquad (5)$$

where $\delta\vec{F}_n^{\mathrm{inter}}$ is the fluctuating part of the intermolecular force acting on segment *n* of the probe macromolecule at t=0, $\hat{L}$ is the Liouville operator of the complete system (probe chain plus matrix), and $\hat{Q}_N = 1 - \hat{P}_N$ where $\hat{P}_N$ is the projection operator onto the phase space of the probe chain (for a detailed description see [31]). The propagator $\exp\left\{i\hat{Q}_N \hat{L} \hat{Q}_N t\right\}$ in expression (5) is called the propagator of the projected/anomalous evolution (see [24, 31]).

The expressions (2)-(4) are not closed, since they reduce the single-chain problem to a problem involving multiple chains, the memory matrix, and the effective potential. This means that converting mathematical relationships into equations requires additional assumptions about the quantities in question.

In our previous work [4] it was suggested to describe the topological interactions between ring polymers in melts in the following way:

$$W\left(\{\vec{r}_i\}\right)=\frac{3k_BT}{2}\sum_n\left(\frac{1}{b^2}\left(\frac{\partial\vec{r}_n}{\partial n}\right)^2+\left(\frac{\vec{r}_n-\vec{r}_{cm}}{\tilde{R}}\right)^2\right) \quad (6)$$

where $\tilde{R}$ is a parameter of the potential. The first term on the right-hand side of Equation (6) describes the conventional effectively intramolecular entropic interactions of an ideal polymer chain [17-21], while the second term, which has the structure of a harmonic potential, is related to the effects of entropic contraction of the polymer chain caused by topological constraints of mutual penetration of cyclic macromolecules into each other.

For the memory matrix, we will use a slightly modified version of the simplest approximation, which can be called the Renormalized Rouse Model approximation (RRM), first described in Ref. [24]. The first approach assumes that the non-entangled Rouse contribution is explicitly isolated from the general relation (2). This is equivalent to assume that the total stochastic force $\vec{F}_n^Q(t)$ includes a $\delta$-function, fast-decaying component $\vec{F}_n^L(t)$. This is a classical stochastic Langevin force, that is local and isotropic, and its time dependence can be approximated by the following way:

$$\left\langle F_n^{L\alpha}(t)F_k^{L\beta}(0)\right\rangle=2\zeta k_BT\delta_{\alpha\beta}\delta_{nk}\delta(t) \quad , \quad (7)$$

where $\zeta$ is the segmental friction coefficient.

After this separation, expression (2) can be rewritten as follows:

$$\frac{\partial}{\partial\vec{r}_n}W\left(\{\vec{r}_i(t)\}\right)=-\zeta\vec{v}_n(t)-\frac{1}{kT}\sum_k\int_0^t d\tau\left\langle\tilde{F}_k^{Q\beta}(0)\tilde{F}_n^{Q\alpha}(\tau)\right\rangle^*_{\gamma_{N(t-\tau)}}v_k^\beta(t-\tau)\vec{e}_\alpha$$
$$+\vec{\tilde{F}}_n^Q(t)+\vec{F}_n^L(t) \quad (8)$$

where $\vec{\tilde{F}}_n^Q(t)$ is the long-living, slowly decaying component of the total stochastic Generalized Langevin force describing dynamical consequences of entanglement effects.

The memory matrix $\Gamma_{nk}^{\alpha\beta}(\tau;t-\tau)=\frac{1}{kT}\left\langle\tilde{F}_k^{Q\beta}(0)\tilde{F}_n^{Q\alpha}(\tau)\right\rangle^*_{\gamma_{N(t-\tau)}}$ in this context is quite general and complex. In principle, this does not rule out reptation movements for linear polymer melts, since the effective averaging $\langle...\rangle^*_{\gamma_{N(t-\tau)}}$ may depend on the conformation of the chain under study in the early stages. However, to the best of our knowledge, the relevant procedure has not yet been implemented with the necessary mathematical details. The next approximation on which RRM is based is pre-averaging, that is, approximating the conditional averaging in the memory matrix using equilibrium averaging [24]. For isotropic polymer systems, expression (8) can then be rewritten as follows:

$$\frac{\partial}{\partial \vec{r}_n} W\left(\{\vec{r}_i(t)\}\right) = -\zeta \vec{v}_n\left(t\right) - \frac{1}{3kT}\sum_k \int_0^t d\tau \left\langle \vec{\tilde{F}}_k^Q(0)\vec{\tilde{F}}_n^Q(\tau)\right\rangle \vec{v}_k(t-\tau) + \vec{\tilde{F}}_n^Q(t) + \vec{F}_n^L(t) \tag{9}$$

In isotropic case any correlations between the mean squared displacements of polymer segments and its initial conformation are lost, and the model becomes dynamically isotropic. The dynamic correlation function $\left\langle \vec{\tilde{F}}_k^Q(0)\vec{\tilde{F}}_n^Q(\tau)\right\rangle$ inside the integral on the right-hand side of expression (9) remains highly complex and requires further approximations. First, it requires knowing the potential associated with intermolecular forces, and second, relating the correlation function of a three-body nature to binary correlation functions. The original article [24] proposed treating the intermolecular potential between polymer segments as a hard-sphere potential. This approximation represents a tedious calculation and is not of fundamental importance, assuming real potential behaves like a hard sphere potential at short distances. The following approximation, proposed in [24], is essentially a modified version of the dynamic superposition approximation, which allows the kernel $\left\langle \vec{\tilde{F}}_k^Q(0)\vec{\tilde{F}}_n^Q(\tau)\right\rangle$ of equation (9) to be represented as the following expansion in terms of wave vectors $\vec{k}$ :

$$\left\langle \vec{\tilde{F}}_k^Q\left(0\right)\cdot \vec{\tilde{F}}_n^Q\left(\tau\right)\right\rangle = \frac{8}{27}\rho_s g^2\left(d^+\right)d^6\left(k_B T\right)^2 \int_0^{b^{-1}} \omega^Q{}_{kn}\left(k,t\right)S^Q\left(k,t\right)k^4 dk \tag{10}$$

where $\rho_s$ is the density of polymer segments in the melt, $d$ is the hard-core diameter of the intermolecular potential, $g\left(d^+\right)$ is the contact value of the intermolecular pair correlation function. . It is convenient to present the correlation functions in Fourier or k-space using Parsevals theorem (see Ref. [24]). The quantity $\omega^Q{}_{kn}\left(k,t\right)$ is the projected evolution of the intramolecular dynamical structure factor and $S^Q\left(k,t\right)$ is the projected evolution of the collective dynamical structure factor. Additional assumptions are required to account for the time dependence of these quantities.

In this paper, we will use the exact result for the projected dynamical evolution in the case where the quantities in the probe chain are purely intramolecular. When the projection operator is a projection onto the phase space of the probing macromolecule, all pure intramolecular quantities remain unchanged over time under the action of the projected propagator (see details in [31], page 56):

$$\begin{aligned}\omega^{Q}{}_{kn}(k,t) &= \left\langle \exp\left\{i\hat{Q}_N\hat{L}\hat{Q}_N t\right\}\exp\left\{i\vec{k}\cdot\left(\vec{r}_k-\vec{r}_n\right)\right\}\right\rangle \\ &= \omega_{kn}(k) \equiv \left\langle \exp\left\{i\vec{k}\cdot\left(\vec{r}_k-\vec{r}_n\right)\right\}\right\rangle\end{aligned} \quad . \tag{11}$$

For ring polymers with central harmonic potential, the right-hand part of expression (11) was already calculated in our previous paper [4].

The projected collective dynamical factor can always be rewritten in the generalized Vineyard form [24]:

$$\hat{S}^{Q}(k,t) \equiv \hat{S}(0)\exp\left\{-\frac{k^2}{6}\left\langle r^2(t)\right\rangle_Q\right\} \tag{12}$$

where $\left\langle r^2(t)\right\rangle_Q$ is an unknown quantity, possibly wave vector dependent, $\hat{S}(0)=\rho_s k_B T\kappa$ is dimensionless compressibility of the melt, $\kappa$ is the isothermal compressibility. As for the new unknown quantity $\left\langle r^2(t)\right\rangle_Q$ , which may be referred to as the projected root-mean-square displacement of the matrix segments, additional assumptions are required. It is worth noting that the Fourier transform of the exponential term on the right-hand side of equation (12), if we neglect the dependence $\left\langle r^2(t)\right\rangle_Q$ on the wave vector, is the propagator of the diffuse dissipation of concentration fluctuations in the matrix over distances of the order of $\left\langle r^2(t)\right\rangle_Q^{1/2}$ during a time interval $t$. We can view this as a known quantity derived from additional reasoning. In any case, it allows us to study a broad class of dynamic polymer models and identify some of their common properties.

The simplest approximation, also used in the initial formulation in Ref. [24], is to neglect the dependence of $\left\langle r^2(t)\right\rangle_Q$ on the wave vector and approximate it using the standard Rouse dynamics.

Let us consider the cyclic Rouse normal modes for ring macromolecules:

$$\vec{X}_p(t) \equiv \frac{1}{N}\sum_{n=0}^{N-1}\vec{r}_n(t)e^{-\frac{2\pi}{N}ipn} \tag{13}$$

where $p = 0, \pm 1, \pm 2, ..., \left[\frac{N}{2}\right]$ is the number of the Rouse normal mode, $\left[\frac{N}{2}\right]$ is the integer part of $\frac{N}{2}$, $N \gg 1$, and ring boundary condition $\vec{r}_0(t) \equiv \vec{r}_N(t)$ is assumed. The relation inverse to (13) is the following:

$$\vec{r}_n(t) = \sum_{p=-[N/2]}^{[N/2]} \vec{X}_p(t) e^{\frac{2\pi i p n}{N}} \tag{14}$$

Note that the coordinates of the vector $\vec{r}_n(t)$ are real numbers, which leads to a relationship between the normal modes $\vec{X}_p^*(t) = \vec{X}_{-p}(t)$ , where $\vec{X}_p^*(t)$ is the complex conjugate of $\vec{X}_p(t)$. When N is an even number, the normal modes $\vec{X}_{N/2}(t)$ and $\vec{X}_{-N/2}(t)$ coincide and should be included only once in the expression (14).

Expression (9) can be rewritten in terms of Rouse normal modes using relations (10)-(14) and considering ring architecture symmetry:

$$\begin{aligned}\dot{\vec{X}}_p(t) &= -w_p \vec{X}_p(t) \\ &-\varphi \int_0^t d\tau \int_0^\infty k^4 dk \omega_p(k) \exp\left\{-\frac{k^2 \langle r^2(\tau)\rangle_Q}{6}\right\} \dot{\vec{X}}_p(t-\tau) + \vec{F}_p^Q(t)\end{aligned}, \tag{15}$$

where

$$\varphi = \left(\frac{2}{3}\right)^3 \frac{\rho_s d^6 b^2 g^2 (d^+) \hat{S}(0)}{3\pi^2 \tau_s} \tag{16}$$

$$\omega_p(k) = \frac{1}{N} \sum_{s,t=0}^{N-1} \left\langle \exp\left\{ i\vec{k}(\vec{r}_s - \vec{r}_t)\right\}\right\rangle = 2\,\mathrm{Re} \sum_{m=0}^{[N/2]} \omega_{m0}(k) \exp\left\{\frac{2\pi i}{N} pm\right\} .$$

The quantity $w_p$ is the Rouse rate for the normal mode with number p without entanglement effects in expression (9), i.e. without the integral part on its right-hand side. It was calculated for a Rouse ring with central harmonic potential in our previous paper [4]:

$$\begin{aligned} w_P &= \frac{4}{\tau_s}\left\{\left(\frac{p}{N}\right)^2 + \frac{1}{(2\pi n_b)^2}\right\} = \frac{4}{\tau_s}\left(\frac{p}{N}\right)^2\left(1 + \left(\frac{p_b}{p}\right)^2\right) \\ &\equiv \frac{4}{\tau_s}\left\{\left(\frac{p}{N}\right)^2 + \left(\frac{b}{2\pi \tilde{R}}\right)^2\right\} \end{aligned} \tag{17}$$

where $\tau_s = \frac{\zeta b^2}{3\pi^2 k_B T}$ is the conventional segmental relaxation time (see, for example [17-20]),

$p_b \equiv \frac{1}{2\pi}\frac{N}{n_b}$ is the parameter reflecting the existence of the central harmonic potential in relation (6) through its connection with the parameter of the potential strength $\tilde{R}$, $n_b \equiv \frac{2R_g^2}{b^2}$.

Based on arguments like those of Flory in the framework of mean-field theory, our recent study [4] demonstrated that

$$\tilde{R}^2 = \beta \frac{N^{4/3}}{\rho_s^{4/3} b^2} \tag{18}$$

with a numerical coefficient $\beta$ of order 1. The squared radius of gyration of a ring with a central harmonic potential for $N \gg 1$ is related to the parameter of the harmonic potential by a very simple relationship:

$$R_g^2 = \frac{1}{2}\tilde{R}b. \tag{19}$$

When we compare this with experimental data for ring-shaped PEO melts with a molecular weight of 96 kDa, we find that $\beta = 2.79 \approx 3$ [4].

Relation (9) has a stochastic integral-differential nature. After multiplying it by $\vec{X}_p^*(0)$ and averaging over all stochastic trajectories using relations (10)-(17), this expression can be written in the following form:

$$\frac{d}{dt}C_p(t) = -w_p C_p(t) - \varphi \int_0^t d\tau \int_0^\infty k^4 dk \omega_p(k) \exp\left\{-\frac{k^2 \left\langle r^2(\tau)\right\rangle_Q}{6}\right\} \frac{d}{dt}C_p(t-\tau) \tag{20}$$

with $C_p(t) \equiv \left\langle \vec{X}_p(0) \cdot \vec{X}_p^*(0) \right\rangle$.

It has previously been shown that [4]:

$$\left\langle \left(\vec{r}_n(0) - \vec{r}_m(0)\right)^2 \right\rangle = 2R_g^2 \left(1 - \exp\left(-\frac{|n-m|b^2}{2R_g^2}\right)\right) \tag{21}$$

for $|n-m| \le \frac{N}{2}$.

Therefore we find for $\omega_p(k)$:

$$\omega_p(k) = 2\,\mathrm{Re}\sum_{m=0}^{[N/2]} \omega_{m0}(k)\exp\left\{\frac{2\pi i}{N}pm\right\}$$
$$= 2\,\mathrm{Re}\sum_{m=0}^{[N/2]} \exp\left\{-\frac{k^2R_g^2}{3}\left(1-\exp\left(-\frac{mb^2}{2R_g^2}\right)\right)\right\}\exp\left\{\frac{2\pi i}{N}pm\right\}. \quad (22)$$

To obtain analytical expressions in closed form, it is convenient to use the following approximation:

$$\left\langle\left(\vec{r}_n(0)-\vec{r}_m(0)\right)^2\right\rangle = \begin{cases} |n-m|b^2 & if \quad |n-m|\le n_b \\ 2R_g^2 & if \quad |n-m|> n_b \end{cases} \quad (23)$$

When p=0, the quantity $\omega_{p=0}(k)$ is known as the coherent static structural factor. This case was already discussed in [4]:

$$\omega_0(k) = (N-2n_b)\exp\left(-\frac{k^2R_g^2}{3}\right)+\frac{12}{k^2b^2}\left(1-\exp\left\{-\frac{k^2R_g^2}{3}\right\}\right). \quad (24)$$

We would like to take this opportunity to correct a typographical error in the numerical coefficient inside the first exponent on the right-hand side of expression (24): the factor 1/3 instead of 1/6 in expression (26) from our previous work [4]. The quantity $n_b \propto N^{2/3} << N$ in the limit N >> 1. Therefore, in the approximation under consideration, quantities proportional to $n_b$ can be neglected compared to N. We have retained it as a symbol of the approximation consistent with the exact value of expression (24) in the limit of small wave vectors, which yields the exact result $\omega_0(0)=N$. The numerical multiplier 2 before $n_b$ indicates the presence of close neighbors on both sides of any segment: to the right and to the left.

In general, using the approximation (23), one can obtain the following expression:

$$\omega_p(k) = N\delta_{p0}\exp\left\{-\frac{k^2b^2n_b}{6}\right\}+\frac{2\left(\frac{b^2k^2}{6}\right)}{\left(\frac{b^2k^2}{6}\right)^2+\left(\frac{2\pi}{N}p\right)^2}$$
$$+2\exp\left\{-\frac{k^2b^2n_b}{6}\right\}\mathrm{Re}\left[\frac{\exp\left\{\frac{2\pi i}{N}pn_b\right\}}{\frac{2\pi ip}{N}-\frac{b^2k^2}{6}}-\frac{\exp\left\{\frac{2\pi i}{N}pn_b\right\}}{\frac{2\pi ip}{N}}\right]. \quad (25)$$

It is striking that the first term on the right-hand side of relation (25) contributes only to the coherent structural factor; that is, it is nonzero only when p=0. We will see that this results in a significant separation of the self-diffusion of the ring's center of mass and its internal modes at large values of $N \gg 1$. It should also be noted that the formal divergence of the second term on the right-hand side of expression (25) at p = 0 is actually eliminated by taking the real part of the corresponding expression.

To keep the analysis simple, it is convenient to study the integro-differential equation (20) using the so-called pseudo-Markov approximation, i.e. the integral on its right-hand side approximates as follows:

$$\begin{aligned}&\int_0^t d\tau \int_0^\infty k^4 dk\omega_p(k)\exp\left\{-\frac{k^2\left\langle r^2(\tau)\right\rangle_Q}{6}\right\}\frac{d}{dt}C_p(t-\tau)\\ &\approx\left(\int_0^t d\tau \int_0^\infty k^4 dk\omega_p(k)\exp\left\{-\frac{k^2\left\langle r^2(\tau)\right\rangle_Q}{6}\right\}\right)\frac{d}{dt}C_p(t)\end{aligned} \tag{26}$$

Following this approximation, the formal solution to the corresponding equation for $p \neq 0$ will be as follows:

$$C_p(t)=\exp\left\{-\int_0^t \frac{w_p d\tau}{1+\hat{\Gamma}_p\left(\tau^{-1}\right)}\right\}C_p(0) \tag{27}$$

where

$$\hat{\Gamma}_p\left(\tau^{-1}\right)=\varphi\int_0^\tau d\tau_1\int_0^\infty k^4 dk\omega_p(k)\exp\left\{-\frac{k^2\left\langle r^2(\tau_1)\right\rangle_Q}{6}\right\}. \tag{28}$$

Note that for $\tau=\infty$, $\hat{\Gamma}_p(0)$ is the Laplace-Fourier transform of $\Gamma_p(\tau)\equiv\varphi\int_0^\infty k^4 dk\omega_p(k)\exp\left\{-\frac{k^2\left\langle r^2(\tau)\right\rangle_Q}{6}\right\}$ at frequency $\omega=0$ , and since we will be interested in its asymptotic behavior at large $\tau$ later on, it is convenient to treat $\hat{\Gamma}_p\left(\tau^{-1}\right)$ as a function of $\tau^{-1}$.

The initial values $C_p(0)$ for the cyclic normal modes of a ring macromolecule with a central harmonic potential were considered in [4], and the result obtained is as follows:

$$C_p(0)\equiv\left\langle \vec{X}_p^*(0)\vec{X}_p(0)\cdot\right\rangle=\frac{1}{\frac{4\pi^2 p^2}{Nb^2}+\frac{N}{n_b^2 b^2}}=\frac{Nb^2}{4\pi^2 p^2\left(1+\left(p_b/p\right)^2\right)}\quad . \tag{29}$$

The value of $\left\langle \vec{X}_p^*(0)\cdot\vec{X}_p(0)\right\rangle$ is correctly defined only for $p\neq 0$. For $p=0$, it diverges, since equilibrium averaging assumes the thermodynamic limit $V\to\infty,\rho_s=const$. Nevertheless, no experimentally measurable quantity depends on the absolute value $C_0(0) \equiv \left\langle \vec{X}_0^*(0) \cdot \vec{X}_0(0) \right\rangle$. For instance, the mean squared displacement of a ring segment can, with the help of relation (14), formally be expressed as follows:

$$g_1(t) \equiv \left\langle \left( \vec{r}_n(t) - \vec{r}_n(0) \right)^2 \right\rangle = \left\langle \left( \sum_{p=-[N/2]}^{[N/2]} \left( \vec{X}_p(t) - \vec{X}_p(0) \right) e^{\frac{2\pi ipn}{N}} \right)^2 \right\rangle \, . \qquad (30)$$
$$= 2\left( C_0(0) - C_0(t) \right) + 4\sum_{p=1}^{[N/2]} \left( C_p(0) - C_p(t) \right) \equiv g_3(t) + g_2(t)$$

To simplify the formulae, we will use the notation commonly employed in works on computer simulations (see, for example, [11]): $g_1(t) \equiv \left\langle \left( \vec{r}_n(t) - \vec{r}_n(0) \right)^2 \right\rangle$ is the segment mean squared displacement in the laboratory coordinate frame, $g_2(t) \equiv 4\sum_{p=1}^{[N/2]} \left( C_p(0) - C_p(t) \right)$ is the segment mean squared displacement in the coordinate system relative to the center of mass of the probe macromolecule, and $g_3(t) \equiv \left\langle \left( \vec{X}_0(t) - \vec{X}_0(0) \right)^2 \right\rangle = 2\left( C_0(0) - C_0(t) \right)$ is the mean squared displacement of the center of mass.

Equation (30) shows that the difference $C_0(0) - C_0(t)$ is important, but not the absolute value $C_0(t) \equiv \left\langle \vec{X}_0(t)\vec{X}_0(0) \right\rangle$. To calculate it, we can formally introduce a weak harmonic potential $\frac{k_B T}{2}\tilde{\varepsilon}\vec{X}_0^2$ acting on the centre of mass of the macromolecule with a force parameter $\varepsilon \to 0$, and then take the limit after formally calculating the corresponding physical quantity. Standard calculations show that in this case we have a definite value for

$$C_0(0) = \left\langle \vec{X}_0 \cdot \vec{X}_0 \right\rangle = \frac{3}{\tilde{\varepsilon}} . \qquad (31)$$

The time evolution of the correlation function $C_0(t) = \langle \vec{X}_0(t) \cdot \vec{X}_0(0) \rangle$, based on relation (20) and using the pseudo-Markov approximation, can be represented in the following way:

$$C_0(t) = C_0(0)\exp\left\{ -\tilde{\varepsilon}\int_0^t \frac{k_B T d\tau}{\zeta^*(\tau)} \right\}, \qquad (32)$$

with

$$\zeta^*(\tau) = N\zeta\left(1+\int_0^\tau \Gamma_0(\tau_1)\,d\tau_1\right). \qquad (33)$$

Note that for calculating $\hat{\Gamma}_0(\tau^{-1}) \equiv \int_0^\tau \Gamma_0(\tau_1)\,d\tau_1$, expression (28) can be used with p=0. Then it is possible to see that for $\tilde{\varepsilon} \to 0$

$$\left\langle \left(\vec{X}_0(t) - \vec{X}_0(0)\right)^2 \right\rangle = 6k_BT\int_0^t \frac{d\tau}{N\zeta\left(1+\int_0^\tau \Gamma_0(\tau_1)\,d\tau_1\right)}. \qquad (34)$$

Now we can rewrite expression (30) as follows:

$$g_1(t) = g_2(t) + g_3(t)$$
$$= 4\sum_{p=1}^{[N/2]} C_p(0)\left(1-\exp\left\{-\int_0^t \frac{w_p\,d\tau}{1+\hat{\Gamma}_p(\tau^{-1})}\right\}\right) + 6\int_0^t \frac{k_BT\,d\tau}{N\zeta\left(1+\int_0^\tau \Gamma_0(\tau_1)\,d\tau_1\right)} \qquad (35)$$

The first term on the right-hand side of expression (35) describes the mean squared displacements in a coordinate system tied to the center of mass, while the second term represents the mean squared displacements of the centers of mass. The globular structure of ring macromolecules is accounted for by the equilibrium values of the Rouse normal modes $C_p(0)$ (Eq. (30)), and the effects of dynamic entanglement are accounted for by the quantities $\hat{\Gamma}_p(\tau^{-1})$ (Eq. (28)). The structure of Eq. (35) allows us to consider the quantities $\hat{\Gamma}_p(\tau^{-1})$ as dimensionless, time-dependent contributions to the friction coefficient associated with the Rouse mode with number $|p| = 0,1,...[N/2]$.

The quantity

$$D_{cm}(t) \equiv \frac{k_BT}{t}\int_0^t \frac{d\tau}{N\zeta\left(1+\int_0^\tau \Gamma_0(\tau_1)\,d\tau_1\right)} = \frac{k_BT}{t}\int_0^t \frac{d\tau}{N\zeta\left(1+\hat{\Gamma}_0(\tau^{-1})\right)} \qquad (36)$$

can be considered as the time-dependent center of mass self-diffusion coefficient.

### 2.2. Renormalized Rouse approximation.

The expressions derived in the previous section of this paper describe a broad class of possible dynamic models for ring melts; the specific details depend on the quantity $\left\langle r^2(t)\right\rangle_Q$ defined by expression (12), *i.e.* the projected mean-squared displacement. To obtain further quantitative results, additional approximation is required relative to it. In the case of linear polymer melts, as a simplest approximation, in the original study [24] it was proposed to approximate $\left\langle r^2(t)\right\rangle_Q$ by dynamics using those of the standard Rouse Model, the corresponding model was named the Renormalized Rouse Model (RRM). We will now discuss a similar approach for melts consisting of cyclic macromolecules. The main difference from the RRM method for linear melts is now that, instead of the classical Rouse model for polymeric rings, we will use the Rouse model for rings modified by a central harmonic potential, as discussed in [4]. This work had shown that the mean squared displacement of a segment of a Rouse ring with a central harmonic potential (indexed by RH) is described by the following relation:

$$\left\langle\left(\vec{r}_n(t)-\vec{r}_n(0)\right)^2\right\rangle^{RH}=\frac{2}{\pi^2}\frac{b^2 t}{N\tau_s}+4\sum_{p=1}^{[N/2]}C_p(0)\left(1-\exp\left\{-w_p t\right\}\right). \tag{37}$$

This relation can also be obtained from (35) by setting $\Gamma_p(\tau)=0$ for all $p=0,1,...,[N/2]$. It has the following asymptotic behavior:

$$\left\langle\left(\vec{r}_n(t)-\vec{r}_n(0)\right)^2\right\rangle^{RH}=\begin{cases}\dfrac{2}{\pi^2}\dfrac{b^2}{N}\dfrac{t}{\tau_s}+\dfrac{2b^2}{\pi^{3/2}}\left(\dfrac{t}{\tau_s}\right)^{1/2} & if \quad \tau_s \ll t \ll \pi^2\tau_s n_b^2 \\[2ex] \dfrac{2}{\pi^2}\dfrac{b^2}{N}\dfrac{t}{\tau_s}+n_b b^2 & if \quad \pi^2\tau_s n_b^2 \ll t\end{cases}. \tag{38}$$

It is convenient to use the following approximation for (38), which exhibits the same asymptotic behavior:

$$\left\langle\left(\vec{r}_n(t)-\vec{r}_n(0)\right)^2\right\rangle^{RH}=\frac{2}{\pi^2}\frac{b^2}{N}\frac{t}{\tau_s}+\frac{\dfrac{2b^2}{\pi^{3/2}}\left(\dfrac{t}{\tau_s}\right)^{1/2}}{1+\dfrac{2}{\pi^{3/2}n_b}\left(\dfrac{t}{\tau_s}\right)^{1/2}}. \tag{39}$$

Given the assumption of RRM that

$$\left\langle r^2(t)\right\rangle_Q=\left\langle\left(\vec{r}_n(t)-\vec{r}_n(0)\right)^2\right\rangle^{RH}, \tag{40}$$

we can derive the general expressions from the previous section.

Let us start the discussion with the self-diffusion *of the center of mass*, i.e. $g_3(t)$. For very short times $t \to 0$ we can neglect $\Gamma_0(\tau_1)$ in expressions (33) and (34). Therefore, neither entanglement nor the effect of the harmonic potential which, in our approach, represents constant topological constraints, is significant. The self-diffusion of the ring center of mass is normal Fickian diffusion and corresponds to the Rouse model prediction which is the same for both linear and ring macromolecules:

$$g_3(t) = \frac{2}{\pi^2} \frac{b^2}{N\tau_s} t + \dots \quad . \tag{41}$$

Then, at longer times, the term in expressions (33) and (34) associated with $\Gamma_0(\tau_1)$ becomes dominant, and we can give the following estimate of the subsequent time dependence for $g_3(t)$, using expressions (24), (25), and (28) (see details of calculations in Appendix):

$$g_3(t) = \frac{1}{2^2 3^{1/2} \pi^{11/4}} \frac{b^2}{\psi N} \left( \frac{t}{\tau_s} \right)^{3/4} + \dots \tag{42}$$

with

$$\psi \equiv \frac{\rho_s d^6 g^2(d) \hat{S}(0)}{b^3} .$$

By comparing expressions (41) and (42), one can determine the characteristic time of transition from the non-entangled Rouse behavior to the entangled one for the diffusion of the center of mass, i.e. from (41) to (42),:

$$\begin{aligned} &\frac{1}{2^2 3^{1/2} \pi^{11/4}} \frac{b^2}{\psi N} \left( \frac{t}{\tau_s} \right)^{3/4} \approx \frac{2}{\pi^2} \frac{b^2}{N\tau_s} t \\ &\tau_{1e} = \left( \frac{\pi^2}{2^3 3^{1/2} \pi^{11/4} \psi} \right)^4 \tau_s = \frac{1}{2^{12} 3^2 \pi^3} \tau_s \psi^{-4} \approx 10^{-6} \frac{\tau_s}{\psi^4} \end{aligned} \quad . \tag{43}$$

On the other hand, by rewriting the characteristic time as $\tau_{1e} \equiv \tau_s \tilde{N}_{1,e}^2$ one can see the following characteristic molecular mass

$$\tilde{N}_{1e} = \frac{1}{192 \pi^{3/2}} \psi^{-2} \approx \frac{3 \cdot 10^{-3}}{\psi^2} . \tag{44}$$

We observe anomalous behavior with the exponent of ¾, as is the case when applying the Renormalize Rouse Model to linear polymer melts (see [24-27]). The only difference here lies in the numerical factor, since in this work we used the exact relation (11) for the projected time evolution of purely intramolecular quantities. The main difference from the linear case is that, for the ring model with a central harmonic potential under consideration, this behavior is now determined by the longest relaxation time of the internal modes $\tau_1^{RH} = \pi^2 n_b^2 \tau_s$, rather than by the Rouse relaxation time, as is the case for linear chains. Another striking feature is the small numerical coefficients in the relations (43) and (44). It results from the fact that if the parameter $\psi$ is such that $\psi > \sqrt{3\cdot 10^{-3}} \approx 0.05$, the predictions of the Rouse model for center-of-mass diffusion (41) should not hold at all, even at times comparable to segmental relaxation times. To the best of our knowledge (see [11], [12] and the literature cited therein), this behavior is characteristic of computer simulations with any molecular masses. Also neutron scattering experiments found subdiffusive centre-of-mass behavior [32,33].

The latter is related to a fact highlighted in [4], according to which, at times of the order of $\tau_1^{RH} = \pi^2 n_b^2 \tau_s$, the contribution to $\left\langle \left( \vec{r}_n(t) - \vec{r}_n(0) \right)^2 \right\rangle^{RH}$ from displacements in internal Rouse relaxation modes have already reached saturation and become equal to $2R_g^2 = b^2 n_b$, whereas the contribution from the motion of the center of mass itself is significantly smaller. This latter contribution becomes comparable to the contribution from internal modes only at significantly longer time scales of the order of

$$\overline{\tau}_1^{RH*} = \frac{3\pi^2}{2} n_b N \tau_s = \frac{1}{2\sqrt{\beta}} \left( \frac{b}{p_l} \right)^{2/3} N^{1/3} \tau_1^{RH} . \tag{45}$$

where $p_l = \rho_s^{-1} b^2$ is the packing length.

This means that over a time interval $\tau_1^{RH} \div \overline{\tau}_1^{RH*}$, the root-mean-square displacement of the ring macromolecules increase rather slowly: by about a factor of 2 - 3.

For larger timescales $t \gg \overline{\tau}_1^{HR*}$ the proposed model predicts a transition to normal diffusion with the self-diffusion coefficient determined by the following equation (see detail in Appendix):

$$D = \frac{1}{3\pi^2} \frac{b^2}{\tau_s N \left( 1 + \int_0^\infty \Gamma_0(\tau_1) d\tau_1 \right)} = \frac{D_R}{\left( 1 + \frac{2^2}{3^{5/2}\pi^{3/2}} \frac{\psi b^2}{\tau_s} \frac{(N - 2n_b)}{n_b^{3/2}} \frac{1}{D_Q} \right)} \tag{46}$$

RRM approximation assumes that $D_Q = D_R = \dfrac{k_B T}{\zeta N}$, then one finds using expressions (18)-(19):

$$D_{RRM} = \frac{b^2}{3\pi^2\tau_s}\frac{1}{N\left(1+\frac{2^2}{3^{5/2}\pi^{3/2}}\psi\frac{N\left(N-2n_b\right)}{\left(n_b\right)^{3/2}}\right)} \quad . \tag{47}$$

$$\xrightarrow{N\to\infty}\frac{b^2}{\pi^{1/2}\psi\tau_s}\frac{3^{3/2}}{4}\left(\sqrt{\beta}N^{2/3}\left(\frac{p_l}{b}\right)^{2/3}\right)^{3/2}\frac{1}{N^3}\propto\frac{b^2}{\psi\tau_s}\left(\frac{p_l}{b}\right)\frac{1}{N^2}$$

Equation (47) shows that there is another characteristic molecular weight:

$$N_{De} = \frac{1}{\psi}\frac{3^{5/2}\pi^{3/2}}{4}\frac{n_b^{3/2}}{N} = \frac{1}{\psi N}\frac{3^{5/2}\pi^{3/2}}{4}\left(\sqrt{\beta}N^{2/3}\left(\frac{p_l}{b}\right)^{2/3}\right)^{3/2} \quad . \tag{48}$$

$$\approx 5\cdot 10^2\psi^{-1}\left(\frac{p_l}{b}\right)$$

At low molecular weights $N < N_{De}$ , we observe Rouse dynamics with respect to the center-of-mass self-diffusion coefficients $D_{RRM} \propto N^{-1}$ , and at higher molecular weights $N > N_{De}$, this dependence becomes significantly stronger $D_{RRM} \propto N^{-2}$ .

Now let us discuss segment mean squared displacements in a coordinate system bounded by the *center of mass* of the tracer chain, i.e., $g_2\left(t\right)$. To do this, it is necessary to use expressions (28) – (31). For short times $t \ll \tau_{1e}$ one naturally obtains the same results as for the classical Rouse model:

$$g_2\left(t\right) = \frac{2b^2}{\pi^{3/2}}\left(\frac{t}{\tau_s}\right)^{1/2} + \ldots \tag{49}$$

For larger time scales, $t \gg \tau_{1e}$ as is the case with $g_3\left(t\right)$ the exponent of the time dependence decreases from 1/2 to 2/5 (use expressions (25), (35) and (38), see also details in Appendix):

$$g_2\left(t\right) = \frac{b^2}{\pi^2}\Gamma\left(\frac{3}{5}\right)\left(\frac{3}{2\pi}\right)^{6/5}\left(\frac{t}{2\psi\tau_s}\right)^{2/5} + \ldots \quad . \tag{50}$$

This behavior will hold up to times of order

$$\tau_{b,1} = \frac{8\pi^2}{9\sqrt{3}}\psi n_b^{5/2}\tau_s \approx \frac{8\pi^2}{3}\psi\left(\frac{p_l}{b}\right)^{5/3}N^{5/3}\tau_s \quad , \tag{51}$$

eventually approaching very slowly its saturation value $g_2\left(\infty\right) = 2R_g^2 = n_b b^2$ .

The terminal relaxation time of the Rouse normal modes for N>>1 is given by the following expression:

$$\tau_1 \equiv \frac{1}{w_p}\left(1+\hat{\Gamma}_1\left(\tau^{-1}\right)\right) = \pi^2\left(1+\int_0^\infty \Gamma_1\left(\tau_1\right)d\tau_1\right)n_b^2\tau_s$$
$$\propto \psi n_b^{3/2} N\tau_s \propto \psi\frac{p_l}{b}\tau_s N^2 \qquad (52)$$

One can see that $\tau_1 \gg \tau_{b,1}$, for large values of $N \gg 1$, this means that over a long period of time $\tau_{b,1} \div \tau_1$, this quantity $g_2\left(t\right)$ will change by approximately a factor of two, reaching its saturation value $n_b b^2$.

It is very important to note that the mean squared displacement of the center of mass during this time is significantly smaller:

$$g_3\left(\tau_1\right) = 6D\tau_1 \propto \frac{b^2}{\psi\tau_s}\left(\frac{p_l}{b}\right)\frac{1}{N^2}\psi\frac{p_l}{b}\tau_s N^2$$
$$\propto b^2\left(\frac{p_l}{b}\right)^2 \qquad (53)$$

and reaches comparable value at essentially longer time:

$$\tau_{1,\mathrm{D}} = \frac{1}{6}\frac{n_b b^2}{D_{RRM}} \propto \psi\left(\frac{b}{p_l}\right)^{1/3} N^{8/3}\tau_s\,. \qquad (54)$$

The 8/3 scaling exponent is clearly observed in computer simulations (see, for example, [35] and the references cited therein) for molecular masses $Z \le 60$. Similarly, a very likely indication of the difference between the time scales $\tau_1 \propto \psi\frac{p_l}{b}\tau_s N^2$ and $\tau_{1,\mathrm{D}} \propto \psi\left(\frac{b}{p_l}\right)^{1/3} N^{8/3}\tau_s$ is the decrease in the effective exponent for $g_1\left(t\right)$ down to 1/4 before reaching the normal diffusion regime in the computer simulation described in the paper [12]( see Fig. 1c).

### 2.3. Some heuristic ideas for expanding the RRM approximation: refined self-consistent approximation.

The RRM described above is the simplest way to approximate the projected dynamics hidden in the quantity $\left\langle r^2\left(t\right)\right\rangle_Q$, which is in fact determined by expression (12). As we noted above (see also detailed discussions in [31]), this problem is one of the fundamental open problems in statistical physics that currently has no exact analytical solution. It is useful to understand that there are no rigorous mathematical theorems that guarantee exact analytical solutions for any problems in statistical physics. In this light, statistical physics is in some respects an art of approximations, the success of which is ultimately determined by their predictive power and agreement with experimental results. We will conclude this section of the article by discussing a heuristic approach to approximating projection dynamics, which, in our view, offers interesting prospects for further research based on the memory matrix formalism. This approach is based on the idea that, for a system of harmonic oscillators, it is possible in principle to obtain a mathematically exact result regarding the nature of projection dynamics.

Let us consider an arbitrary set of coupled harmonic oscillators. We can consider a subset of these oscillators as a probe macromolecule, and the remainder as the matrix. For this probe macromolecule, we can derive equation (1) using the projection operator on the phase space of the probe macromolecule. In this case, it can be rigorously proven that the projected dynamics coincide exactly with the Hamiltonian dynamics of the system under consideration, provided that all the oscillators of the macromolecule probe are fixed. Formally, this is equivalent to subtracting the kinetic energy of a set of test oscillators from the total Hamiltonian of the entire system (for details, see [31], pp. 60–62).

We will try to use this fact for formulating a new approximation to the self-diffusion coefficient. Let us to return to the last line of expression (46):

$$D = \frac{D_R}{\left(1 + \tilde{\psi} D_Q^{-1}\right)} \tag{55}$$

with $\tilde{\psi} = \frac{2^2}{3^{5/2}\pi^{3/2}} \frac{\psi b^2}{\tau_s} \frac{\left(N - 2n_b\right)}{n_b^{3/2}}$.

The quantity $D_Q$ represents the self-diffusion coefficient of neighboring macromolecules around the probe macromolecule. If the probe macromolecule is immobile, this means that it does not affect the friction coefficient of the neighboring chains and exerts only a static force. Therefore, it seems natural to assume that $D_Q$ is slightly larger than $D$. This difference can be expressed in terms of the dependence of the self-diffusion coefficient on molecular weight:

$$D_Q = D\left(N\left(1-\varepsilon\right)\right) \tag{56}$$

where $\varepsilon$ is some small numerical parameter considering immobile behaviour of the probe chain on projected dynamics. For large molecular masses $D_Q\left(N\right) \to 0$ , we can approximate equation (55) as follows:

$$D\left(N\right) = \frac{1}{\tilde{\psi}} D_R D_Q\left(N\right) = \frac{1}{\tilde{\psi}} D_R D\left(N\left(1-\varepsilon\right)\right) \tag{57}$$

The functional equation (57) and neglecting the N dependence of $\tilde{\psi}$ and $\varepsilon$, has an exact solution:

$$D(N)=\frac{\tilde{D}}{\sqrt{N}}\exp\left\{\frac{\ln(N)\cdot\ln\left(N\left(\frac{\tilde{\psi}}{D_0}\right)^2\right)}{2\ln(1-\varepsilon)}\right\}, \tag{58}$$

where $\tilde{D}$ is a constant, $D_0=\frac{b^2}{3\pi^2\tau_s}$ is the so called segmental diffusion coefficient. Correctness of the expression (58) can be checked by direct substitution into (57). For large $N\gg\left(\frac{\tilde{\psi}}{D_0}\right)^2$ and small $\varepsilon\ll 1$ the expression (58) can be simplified:

$$D(N)=\frac{\tilde{D}}{\sqrt{N}}\exp\left\{-\frac{\ln^2(N)}{2\varepsilon}\right\} \quad . \tag{59}$$

Expressions (58) and (59) describe a super-polynomial decay of the self-diffusion coefficient, i.e. faster than $N^{-m}$ for any $m$ as the molecular mass increases. This behavior can be viewed as a kind of dynamic localization. Qualitatively, this prediction can be viewed as analogous to the dynamic localization predicted in [7, 9, 10] based on, at first glance, different arguments and with a stronger dependence on molecular mass $D(N)\propto\exp\{-cN\}$, where $c$ is a constant.

It seems instructive also to discuss the limiting case, when $\varepsilon=0$, which can be regarded as a naive self-consistent approximation, the original equation (55) admits an exact solution without additional approximations:

$$D(N)=D_R-\tilde{\psi}\,. \tag{60}$$

The self-diffusion coefficient cannot be negative; a self-consistent approximation predicts that $D(N)=0$ for molecular masses $N\geq\tilde{N}$, where $\tilde{N}\propto\frac{b}{p_l}\psi^{-1}$ satisfy the equation:

$$D(\tilde{N})=D_R(\tilde{N})-\tilde{\psi}=0\,. \tag{61}$$

This kind of prediction can be named as the strong localization, actually this means absents of transition to normal diffusion at any times at all, i.e. the mean squared displacement of the center of mass will increase slower, that time $g_3(t)\propto t^{\gamma}$ , $\gamma<1$.

It should also be noted that results (58) and (59) follow from the assumption that the quantity $\varepsilon$ which, at this stage of the analysis, is regarded as a measure of the decrease in the local friction coefficient of the surrounding macromolecules of the matrix due to the immobility of the test macromolecule, does not depend on N. If the parameter $\varepsilon$ decreases as N increases, dynamic localization will remain, and the molecular-mass dependence of the self-diffusion coefficient will lie between the relationships given by (59) and (61).The main objective of this section of the article is to demonstrate, based on the memory function formalism, the fundamental possibility of dynamic localization, that is, a decrease in the self-diffusion coefficient with increasing molecular mass that is faster than any power law. At this stage of our understanding of the problem, we do not claim to have a definitive quantitative solution. Here we are dealing with an open question directly related to the connection between projection dynamics and real dynamics, which requires further investigation and will be the subject of our future research.

## 3. Discussion and conclusions.

In this paper, we have perhaps made the simplest extension of our previous work [4] to describe the main features of ring polymer melts using dynamic equations of motion linked to the fundamental results of general statistical mechanics. The entanglement phenomena in the system under consideration can be divided into two categories: topological effects/interactions, which do not disappear over time, and dynamic effects, which disappear after a time exceeding the finite relaxation time. In our approach, topological effects are described by the effective central harmonic potential having an entropic nature, introduced in the second term on the right-hand side of equation (6). The effects associated with the dynamic entanglements are described by the memory matrix in equation (2). Approximating the time evolution of the memory function using standard Rouse dynamics, that is, using equation (39), allows us to construct a mathematically self-consistent model that exhibits rich behavior closely resembling that observed experimentally and in numerical simulations [1].

First and foremost, there is a clear distinction between the relaxation processes of the internal normal modes and the self-diffusion of the ring's center of mass. The terminal relaxation time for Rouse internal normal modes is given by the expression:

$$\tau_1 \equiv \frac{\hat{\Gamma}_1(0)}{w_1} \approx 3\pi\beta^{3/4}\left(\frac{p_l}{b}\right)^{2/3}\psi N^2\tau_s . \qquad (62)$$

The displacement of the ring's center of mass over time scales on the order of $\tau_1$ is significantly smaller than its linear dimension $R_g = \frac{b n_b^{1/2}}{\sqrt{2}}$ and reaches the same order of magnitude only over much longer time scales:

$$\tau_{1,\mathrm{D}} = \frac{1}{6}\frac{n_b b^2}{D_{RRM}} \propto \psi \left(\frac{p_l}{b}\right)^{-1/3} N^{8/3}\tau_s \,. \qquad (63)$$

The self-diffusion coefficient of the molecular mass for the Renormalized Rouse ring model with a central harmonic potential is, for sufficiently large molecular masses, given by:

$$D_{RRM} \propto \frac{b^2}{\psi \tau_s}\left(\frac{p_l}{b}\right)\frac{1}{N^2} \,. \qquad (64)$$

In essence, this power law dependence on the mass,  is more pronounced than in the case of linear polymer melts Renormalized Rouse model, where renormalization described above yields the result $D_{RR} \propto N^{-3/2}$ (see [24,26–29]).  We observe that the internal harmonic potential and the memory function enhance the mutual influence on the self-diffusion coefficient. A related phenomenon was reported in a recent publication [34], which discussed the behavior of a linear Rouse chain modified by adding by a phenomenological memory function in an external harmonic potential. In our case the harmonic potential is internal, the chain is cyclic and for memory we used kernel derived based on Renormalized Rouse approach.  Note also that the molecular mass dependence predicted by relation (61) is very close to that observed experimentally and in simulations for ring melts with $Z \le 60$, where $Z = N / N_e$, $N_e$ is the number of Kuhn segments in linear polymer melts between neighboring entanglements (see, for example, [12]).  In the present model, the predictions for times shorter than $t \ll \tau_{1,\mathrm{D}}$ are also in good agreement with experimental and simulation data for ring melts with $Z \le 60$.  The anomalous exponents for the time dependence of segment and center of mass mean squared displacements for linear chains given by expressions (42), (50) are well known and can be found in those references [24,26–29].  In particular, ring size dependence of the center-of-mass diffusion appearing in the present model is $N^{-2}$. Experiments find that the Fickian diffusion coefficients as a function of ring size follow a very similar power law as that for their linear counterparts $N^{-2.2\text{-}2.3}$ which was found also in good agreement with simulation and the topological free volume theory of Sakaue (see in [36]). However, the result disagrees with the predictions from the scaling models that always predict power law exponents less than 2. Centre-of-mass MSD in the time interval covering the distances comparable with the characteristic ring size demonstrates sub-diffusive behavior $t^{3/4}$ which was observed by neutron scattering experiments.  The molecular mass dependence of the segmental MSD $N^{-8/3}$ we found is also observed in computer simulations (see, for example, [35] and the references cited therein) for molecular masses. Note that all described predictions are the mathematical consequences of the simple approximation (40) for the projected dynamics describing the time dependence of the memory matrix kernel of the dynamical equations (1) and (2) with (6).

For higher molecular weights, as reported in recent studies [7–10], the situation changes, and melts of ring polymers begin to exhibit a transition to a state of dynamic localization. As shown at the end of the previous section of this article, such behavior can in principle be considered within the framework of the memory matrix formalism. If we assume that, for large molecular masses, the refined self-consistent approximation for projected dynamics proves to be more applicable, then the decrease in the self-diffusion coefficient as a function of molecular mass (see expression (58)) becomes, as the molecular mass increases, more rapidly varying than any polynomial law; in other words, we observe a case of dynamic localization. We think that further research in this area holds great promise, and it is and will remain the focus of our future research.

**Supplementary Material.**

A fairly brief and thorough explanation of the Zwanzig–Mori projection operator method, which is necessary for understanding this article, is provided in [31].

**Acknowledgements**

Financial support for this work by DFG under KR 3929/3-1, STA 511/18-1, project number 571095035, is gratefully acknowledged. We are also grateful to the second reviewer for his critical analysis of Section 2.3 of the article and for suggesting solution (58) to us.

# Appendix

## 1. Derivation of expression (42)

By definition, see expression (34), we have:

$$g_3(t) = 6k_BT\int_0^t \frac{d\tau}{\zeta^*(\tau)}. \tag{A1}$$

Then using (33) we can rewrite the expression (A1) as follows:

$$g_3(t) = 6\int_0^t \frac{k_BTd\tau}{N\zeta\left(1+\int_0^\tau \Gamma_0(\tau_1)d\tau_1\right)}. \tag{A2}$$

The quantity $\Gamma_0(\tau)$ can be calculated in the following way:

$$\int_0^\tau \Gamma_0(\tau_1)d\tau_1 = \varphi\int_0^\tau d\tau_1\int_0^\infty \omega_0(k)\exp\left\{-\frac{k^2}{6}\left\langle r_n^2(\tau_1)\right\rangle_Q\right\}k^4dk$$
$$= 27\sqrt{\frac{3\pi}{2}}\varphi\int_0^\tau d\tau_1\left\{\frac{(N-2n_b)}{\left(n_bb^2+\left\langle r_n^2(\tau_1)\right\rangle_Q\right)^{5/2}}+\frac{4}{b^2}\left(\frac{1}{\left\langle r_n^2(\tau_1)\right\rangle_Q^{3/2}}-\frac{1}{\left(n_bb^2+\left\langle r_n^2(\tau_1)\right\rangle_Q\right)^{3/2}}\right)\right\}. \tag{A3}$$

At short times $\tau \ll \tau_s n_b^2$ dominates the term connected with contribution from $\left\langle r_n^2(\tau_1)\right\rangle_Q^{-3/2}$, which for RRM is:

$$\int_0^\tau \Gamma_0(\tau_1)d\tau_1 \simeq 27\sqrt{\frac{3\pi}{2}}\varphi\frac{4}{b^2}\int_0^{\tau_1}\frac{d\tau_1}{\left\langle r_n^2(\tau_1)\right\rangle_Q^{3/2}}$$
$$= \left(\frac{2}{3}\right)^3\frac{\rho_s d^6b^2g^2(d)\hat{S}(0)}{3\pi^2\tau_s}27\sqrt{\frac{3\pi}{2}}\left(\frac{\pi^{3/2}}{2}\right)^{3/2}\frac{16}{b^5}\tau_s^{3/4}\tau^{1/4}. \tag{A4}$$
$$= \frac{32\pi^{3/4}}{\sqrt{3}}\frac{\rho_s d^6g^2(d)\hat{S}(0)}{b^3\tau_s^{1/4}}\tau^{1/4} = \frac{2^5\pi^{3/4}}{3^{1/2}}\psi\left(\frac{t}{\tau_s}\right)^{1/4}$$

Then substituting the expression (A4) into (A2) one can get for large N:

$$g_3(t) = 6\int_0^t \frac{k_BTd\tau}{N\zeta\left(1+\int_0^\tau \Gamma_0(\tau_1)d\tau_1\right)}$$
$$= 6\frac{b^2}{3\pi^2N\tau_s}\int_0^t \frac{d\tau}{1+\frac{2^5\pi^{3/4}}{3^{1/2}}\psi\left(\frac{\tau}{\tau_s}\right)^{1/4}} \approx \frac{1}{2^23^{1/2}\pi^{11/4}}\frac{b^2}{\psi N}\left(\frac{t}{\tau_s}\right)^{3/4}. \tag{A5}$$

## 2. Derivation of expression (46)

The starting point is the expression:

$$D=\lim_{t\to\infty}D(t)\equiv\frac{1}{t}\int_0^t\frac{k_BT}{\zeta^*(t)}=\frac{1}{3\pi^2}\frac{b^2}{\tau_s}\frac{1}{t}\int_0^t\frac{d\tau}{N\left(1+\int_0^{\tau}\Gamma_0(\tau_1)d\tau_1\right)}. \qquad \text{(A6)}$$

At large $\tau\gg\frac{n_bb^2}{6D_Q}\equiv\tau_Q$ the memory function $\Gamma_0(\tau)$ is fast decaying, because it is assumed that

$\left\langle r_n^2(\tau)\right\rangle_Q=6D_Qt+n_bb^2$. (A7) Therefore

the integral $\int_0^{\infty}\Gamma_0(\tau_1)d\tau_1$ exists. We can then take arbitrary $m\gg1$ and rewrite (A6) by the following way:

$$D=\lim_{t\to\infty}\frac{1}{3\pi^2}\frac{b^2}{\tau_s}\frac{1}{t}\left(\int_{m\tau_b}^{t}\frac{d\tau}{N\left(1+\int_0^{\tau}\Gamma_0(\tau_1)d\tau_1\right)}+\int_0^{m\tau_b}\frac{d\tau}{N\left(1+\int_0^{\tau}\Gamma_0(\tau_1)d\tau_1\right)}\right). \qquad \text{(A8)}$$

One can then chose m sufficiently large, for example $m=10^{10}$, and make the following approximation:

$$D=\lim_{t\to\infty}\frac{1}{3\pi^2}\frac{b^2}{\tau_s}\frac{1}{t}\left(\frac{t-m\tau_b}{N\left(1+\int_0^{\infty}\Gamma_0(\tau_1)d\tau_1\right)}+\int_0^{m\tau_b}\frac{d\tau}{N\left(1+\int_0^{\tau}\Gamma_0(\tau_1)d\tau_1\right)}\right). \qquad \text{(A9)}$$

$$=\frac{1}{3\pi^2}\frac{b^2}{\tau_s}\frac{1}{N\left(1+\int_0^{\infty}\Gamma_0(\tau_1)d\tau_1\right)}$$

Now let us discuss the quantity

$$\hat{\Gamma}(0)=\int_0^{\infty}\Gamma_0(\tau)d\tau$$

$$=\frac{2^{5/2}}{3^{1/2}\pi^{3/2}}\frac{\psi b^5}{\tau_s}\int_0^{\infty}d\tau\left\{\frac{(N-2n_b)}{\left(n_bb^2+\left\langle r_n^2(\tau)\right\rangle_Q\right)^{5/2}}+\frac{4}{b^2}\left(\frac{1}{\left\langle r_n^2(\tau)\right\rangle_Q^{3/2}}-\frac{1}{\left(n_bb^2+\left\langle r_n^2(\tau)\right\rangle_Q\right)^{3/2}}\right)\right\}. \qquad \text{(A10)}$$

The main contribution to the mentioned integral comes at times of order $\tau_Q = \frac{n_b b^2}{6D_Q}$, when $\left\langle r_n^2(\tau)\right\rangle_Q \propto n_b b^2$. For $N \gg 1$, $n_b \propto N^{2/3}$ and the contribution from terms inside the internal bracket in the right-hand part of (A10) can be neglected:

$$\hat{\Gamma}(0) = 27\sqrt{\frac{3\pi}{2}}\varphi\int_0^\infty d\tau \frac{(N-2n_b)}{\left(n_b b^2 + \left\langle r_n^2(\tau)\right\rangle_Q\right)^{5/2}}. \quad \text{(A11)}$$

The value $\left\langle r_n^2(\tau)\right\rangle_Q$ in the time range of interest can be estimated as:

$$6D_Q t \le \left\langle r_n^2(\tau)\right\rangle_Q \le n_b b^2 + 6D_Q t. \quad \text{(A12)}$$

From this we can establish the following limitations:

$$\begin{gathered} 27\sqrt{\frac{3\pi}{2}}\varphi\int_0^\infty d\tau \frac{N-2n_b}{\left(2n_b b^2 + 6D_Q\tau\right)^{5/2}} \le \hat{\Gamma}(0) \le 27\sqrt{\frac{3\pi}{2}}\varphi\int_0^\infty d\tau \frac{N-2n_b}{\left(n_b b^2 + 6D_Q\tau\right)^{5/2}} \\ \frac{2b^2}{3^{5/2}\pi^{3/2}D_Q\tau_s}\psi\frac{N-2n_b}{n_b^{3/2}} \le \hat{\Gamma}(0) \le \frac{2^{5/2}b^2}{3^{5/2}\pi^{3/2}D_Q\tau_s}\psi\frac{N-2n_b}{n_b^{3/2}} \end{gathered}. \quad \text{(A13)}$$

We see that lower and upper bounds in expression (A13) differ from each other only by a numerical coefficient $2^{-3/2}$. For simplification in this paper, we are using as a reasonable approximation for the lower bound:

$$\hat{\Gamma}(0) = \frac{2^2}{3^{5/2}\pi^{3/2}}\frac{b^2\psi}{D_Q\tau_s}\frac{N-2n_b}{n_b^{3/2}}. \quad \text{(A14)}$$

and finally:

$$D = \frac{D_R}{\left(1 + \frac{2^2}{3^{5/2}\pi^{3/2}}\frac{\psi b^2}{\tau_s}\frac{(N-2n_b)}{n_b^{3/2}}\frac{1}{D_Q}\right)}. \quad \text{(A15)}$$

**Derivation of expression (49)**

The starting point is expression (35), in which at interesting time interval $\tau \ll \tau_s n_b^2$ dominates the contribution from the second term in the right-hand side:

$$g_1(t) \approx g_2(t) = 4\sum_{p=1}^{[N/2]} C_p(0)\left(1 - \exp\left\{-\int_0^t \frac{w_p d\tau}{1 + \hat{\Gamma}_p\left(\tau^{-1}\right)}\right\}\right). \quad \text{(A16)}$$

Here we have contributions from numerous internal modes of the ring macromolecule, most of which decay by time t, which allows us to use the Markov approximation, i.e. to set $\hat{\Gamma}_p\left(\tau^{-1}\right)=\hat{\Gamma}_p\left(0\right)$:

$$g_2\left(t\right)=4\sum_{p=1}^{[N/2]}C_p\left(0\right)\left(1-\exp\left\{-\frac{w_p t}{1+\hat{\Gamma}_p\left(0\right)}\right\}\right)\ . \tag{A17}$$

If one considers the dynamic polymer models for which the projected mean squared segment displacements power law depends on time:

$$\left\langle r^2\left(t\right)\right\rangle_Q=\left\langle r^2\left(\tau_1^Q\right)\right\rangle_Q\left(\frac{t}{\tau_1^Q}\right)^{\bar{\alpha}}, \tag{A18}$$

where $\left\langle r^2\left(\tau_1^Q\right)\right\rangle_Q$, $\tau_1^Q$ are some constants and exponent $\bar{\alpha}>\frac{2}{7}$, the expression (A17) can be calculated asymptotically exactly in the limit of large $N\gg 1$.

For calculating $\hat{\Gamma}_p\left(0\right)$ let us return to the expression (28). For discussing times $\tau\ll\tau_s n_b^2$ the main contribution comes from the second term in the right part of the expression (25):

$$\hat{\Gamma}_p\left(0\right)=\int_0^\infty\Gamma_p\left(\tau\right)d\tau\cong\varphi\int_0^\infty dt\int_0^\infty k^4dk\frac{2\left(\frac{b^2k^2}{6}\right)}{\left(\frac{b^2k^2}{6}\right)^2+\left(\frac{2\pi}{N}p\right)^2}\exp\left\{-\frac{k^2\left\langle r^2\left(\tau\right)\right\rangle_Q}{6}\right\}. \tag{A19}$$

Then using expression (A18) and changing order of integrations one can get:

$$\hat{\Gamma}_p\left(0\right)=\varphi\int_0^\infty dt\int_0^\infty dkk^4\frac{2\left(\frac{b^2k^2}{6}\right)}{\left(\frac{b^2k^2}{6}\right)^2+\left(\frac{2\pi}{N}p\right)^2}\exp\left\{-\frac{k^2\left\langle r^2\left(\tau_1^Q\right)\right\rangle_Q\left(\frac{t}{\tau_1^Q}\right)^{\bar{\alpha}}}{6}\right\}$$

$$=\frac{1}{\bar{\alpha}b^2}\frac{72\sqrt{6}\varphi\tau_1^Q}{\left\langle r^2\left(\tau_1^Q\right)\right\rangle_Q^{3/2}}\left(\frac{Nb^2}{2\pi p\left\langle r^2\left(\tau_1^Q\right)\right\rangle_Q}\right)^{\frac{1}{\bar{\alpha}}-\frac{3}{2}}\left(\int_0^\infty q^{\frac{2}{\bar{\alpha}}-1}e^{-q^2}dq\right)\left(\int_0^\infty\frac{z^{\frac{1}{\bar{\alpha}}-\frac{5}{2}}}{1+z^2}dz\right)\quad , \tag{A20}$$

$$=A\left(\frac{N}{2\pi p}\right)^{\frac{1}{\bar{\alpha}}-\frac{3}{2}}$$

with

$$A \equiv \frac{1}{\bar{\alpha}} \frac{32}{9\pi^2} \frac{\sqrt{6}\pi\Gamma\left(\frac{2}{\bar{\alpha}}\right)}{\cos\left(\frac{\pi}{2}\left(\frac{5}{2}-\frac{1}{\bar{\alpha}}\right)\right)} \frac{\tau_1^Q}{\tau_s} \left(\frac{b^2}{\left\langle r^2\left(\tau_1^Q\right)\right\rangle_Q}\right)^{\frac{1}{\bar{\alpha}}} \psi \,. \tag{A21}$$

Then it is necessary to return to the expressions (A17) and (29). For discussing times $\tau \ll \tau_s n_b^2$ only the Rouse modes with $p \geq p_b$ are important. Then we can calculate by the following way:

$$\begin{aligned} g_2(t) &\approx 4\sum_{p=p_b}^{N/2} \frac{Nb^2}{4\pi^2 p^2}\left(1-\exp\left\{-\frac{4}{\tau_s}\left(\frac{p}{N}\right)^{\frac{1}{\bar{\alpha}}+\frac{1}{2}} \frac{(2\pi)^{\frac{1}{\bar{\alpha}}-\frac{3}{2}}}{A} t\right\}\right) \\ &\approx \int_0^\infty dp \frac{Nb^2}{\pi^2 p^2}\left(1-\exp\left\{-\frac{4}{\tau_s}\left(\frac{p}{N}\right)^{\frac{1}{\bar{\alpha}}+\frac{1}{2}} \frac{(2\pi)^{\frac{1}{\bar{\alpha}}-\frac{3}{2}}}{A} t\right\}\right) \\ &= \frac{b^2}{\pi^2}\Gamma\left(\frac{1-\bar{\alpha}}{2+\bar{\alpha}}\right)\left(\frac{4(2\pi)^{\frac{1}{\bar{\alpha}}-\frac{3}{2}}}{A}\frac{t}{\tau_s}\right)^{\frac{2\bar{\alpha}}{\bar{\alpha}+2}} \end{aligned} \quad . \tag{A22}$$

For the RRM approximation for times $\tau \ll \tau_s n_b^2$

$$\left\langle r^2(t)\right\rangle_Q = \frac{2b^2}{\pi^{3/2}}\left(\frac{t}{\tau_s}\right)^{1/2}, \tag{A23}$$

and $\bar{\alpha} = 1/2$, therefore

$$\begin{aligned} g_2(t) &\approx 4\sum_{p=p_b}^{N/2} \frac{Nb^2}{4\pi^2 p^2}\left(1-\exp\left\{-\frac{4}{\tau_s}\left(\frac{p}{N}\right)^{\frac{1}{\bar{\alpha}}+\frac{1}{2}} \frac{(2\pi)^{\frac{1}{\bar{\alpha}}-\frac{3}{2}}}{A} t\right\}\right) \\ &\approx \frac{1}{\pi^2}\int_0^\infty dp \frac{Nb^2}{p^2}\left(1-\exp\left\{-\frac{4}{\tau_s}\left(\frac{p}{N}\right)^{\frac{1}{\bar{\alpha}}+\frac{1}{2}} \frac{(2\pi)^{\frac{1}{\bar{\alpha}}-\frac{3}{2}}}{A} t\right\}\right) \\ &= \frac{1}{\pi^2}\Gamma\left(\frac{3}{5}\right)\left(\frac{3}{2\pi}\right)^{6/5}\left(\frac{t}{2\psi\tau_s}\right)^{2/5} b^2 \end{aligned} \quad . \tag{A24}$$